# Deep Learning for Sleep Stages Classification: Modified Rectified Linear Unit Activation Function and Modified Orthogonal Weight Initialisation

Akriti Bhusal [1], Abeer Alsadoon [1,2]*, P.W.C. Prasad [1], Nada Alsalami [3], Tarik A. Rashid[4]

[1]School of Computing and Mathematics, Charles Sturt University, Sydney, Australia.
[2]Asia Pacific International College (APIC), Information System Department, Sydney, Australia
[3]Computer Science Department, Worcester State University, MA, USA
[4]Computer Science and Engineering, University of Kurdistan Hewler, Erbil, KRG, IRAQ

**Abeer Alsadoon[1*]**

* Corresponding author. A/Prof (Dr) Abeer Alsadoon, [1]School of Computing and Mathematics, Charles Sturt University, Sydney, Australia. Email: alsadoon.abeer@gmail.com , Phone +61 413971627

## Abstract

**Background and Aim** Each stage of sleep can affect human health, and not getting enough sleep at any stage may lead to sleep disorder like parasomnia, apnea, insomnia, etc. Sleep-related diseases could be diagnosed using Convolutional Neural Network Classifier. However, this classifier has not been successfully implemented into sleep stage classification systems due to high complexity and low accuracy of classification. The aim of this research is to increase the accuracy and reduce the learning time of Convolutional Neural Network Classifier. **Methodology** The proposed system used a modified Orthogonal Convolutional Neural Network and a modified Adam optimisation technique to improve the sleep stage classification accuracy and reduce the gradient saturation problem that occurs due to sigmoid activation function. The proposed system uses Leaky Rectified Linear Unit (ReLU) instead of sigmoid activation function as an activation function. **Results** The proposed system called Enhanced Sleep Stage Classification system (ESSC) used six different databases for training and testing the proposed model on the different sleep stages. These databases are University College Dublin database (UCD), Beth Israel Deaconess Medical Center MIT database (MIT-BIH), Sleep European Data Format (EDF), Sleep EDF Extended, Montreal Archive of Sleep Studies (MASS), and Sleep Heart Health Study (SHHS). Our results show that the gradient saturation problem does not exist anymore. The modified Adam optimiser helps to reduce the noise which in turn result in faster convergence time. **Conclusion** The convergence speed of ESSC is increased along with better classification accuracy compared to the state of art solution.

# 1. Introduction

Sleep Stage Classification is traditionally conducted by Polysomnography Scoring (PSG). Patients have to spend a night in the PSG recording centres and attach complex equipment in order to record the sleep signals which is inconvenient hence decreases the accuracy of reading due to sleep obstructions. The manual scoring of PSG signals is labour intensive and time-consuming [1]. Event scoring is prone to fatigue errors due to its monotonous and laborious nature [2]. Also, intra and inter-rater reliability can affect classification. Due to this, sleep stage classification suffers from substantial variability which reduces the accuracy of the process [1]. Sleep Stage Classification in biomedical engineering has been greatly enhanced by Deep Learning, which is a relatively new technology Deep learning and Convolutional Neural Network (CNN) is used to increase the accuracy of automatic sleep stage classification and reduce the learning time of the model. CNN has the capacity to extract high-level features from each new layer using the information from the input signal. Such models generally fail to extract a rich and a diverse feature from the signal. The current CNN systems are not compatible with the portable wearable devices with low computational power. With the use of orthogonally initialised weights, Orthogonal Convolutional Neural Network (OCNN) makes it possible for this model to run on portable devices and allow the patient to record sleep data from the place of their convenience without any sleep obstructions [2][3].

For accurate sleep stage classification, it is highly desirable for the network model to provide an accurate prediction within the least possible training time. Technically, Hilbert Huang transform converts EEG signal into a Time-Frequency Image representation which better describes the EEG signal for a CNN and allows it to extract more rich and diverse features. Zhang et al. [4] used the Sigmoid function as an activation function. This work suffers from gradient saturation which causes the weights to be updated slowly, hence reducing the accuracy of classification [4] [5]. In order to overcome this, Mousav et al. [5] utilise the ReLU function as an activation function to eliminate the problem of gradient saturation and hence increase the accuracy. Figure 1 shows three different classification systems.

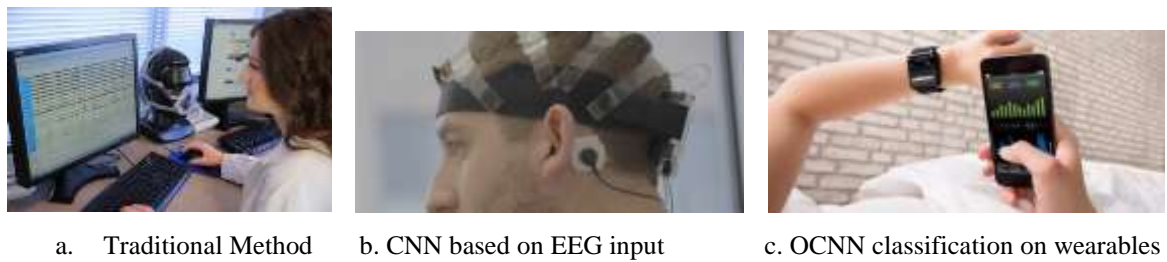

a. Traditional Method     b. CNN based on EEG input     c. OCNN classification on wearables

Figure 1 - (a) Traditional Method (b) CNN based on EEG input (c) OCNN classification on wearables
[These images are downloaded from Google Search Engine]

The aim of our research is to propose an OCNN model based on a ReLU activation function to remove the problem of gradient saturation. Furthermore, this research aims to improve the classification accuracy by removing the problem of slow weight updates caused by gradient saturation in a sigmoid function. This study proposes a modified Leaky ReLU activation function for the OCNN model along with the addition of Adam optimiser [6] to increase the accuracy of classification and reduction of training time, i.e. processing time of the model.

This paper is organized as: In section 2, a literature review is presented with the state of art explanation. In section 3, the proposed system is explained with its improvement over the state of art solution. In section 4, the results of the proposed system are given. The results are discussed and compared with the other existing solution in section 5. Finally, we conclude the study in the section 6.

## 2. Literature Review

As given in the rules of Rechtschaffen and Kales (R&K), a sleep stage is classified as Wake (W), 4 stages from S1 to S4 for non-rapid eye movement (NREM), and rapid eye movement (REM). S3 and S4 are replaced by a single class as slow-wave sleep (SWS). That means S1 from NREM is the first stage of a sleep, S2 is the stage when the sleep begins without eye movements. In REM stage, eyes are closed, however they are moving fast [2][5].

To understand the concept of neural network models for Sleep Stage Classification for detection of sleep stages and sleep related disorders, we studied relevant papers in the field which are analysed in the following subsections. These papers are organized into three classes according to their type: Extracting features from the input signal, Neural Network Models for sleep Stage Classification, and the state of art paper.

### 2.1 Extracting Features from the input signal

Phan et al. [7] enhanced the accuracy of sleep stage classification for each signal epoch and the accuracy of prediction of sleep stage for its neighbouring epochs. This has been done by producing multiple decisions on the same epoch. This method utilises the strong dependency of the class in neighbouring epochs to reduce the processing time and enhance performance. It provides an accuracy of 82.3% and 83.6% for Sleep EDF Expanded (Sleep EDF) dataset and Montreal Archive of Sleep Studies (MASS) respectively. However, with a large output context, the link between neighbouring epochs and the input epoch weakens, which affects the reliability of the result that is calculated by aggregation [7].

Xu et al.[8] and Zhang et al.[1] offered a solution using the temporal information to improve the accuracy of classification of sleep stages with time-frequency spectra from 30s epochs. Zhang at al. [1] compared two different feature extraction methods, one using a raw PSG signal as input and the second using a Short Time Fourier Transform (STFT) spectrograms of the signal. This work uses a Tukey window to help in examining proximal and sequential relationships of the frequencies in time and epoch. Xu's research resulted in the weighted F1-score of 0.87 and Cohen's Kappa(K) of 0.82.  Xu et al. [4] proposed a CNN model with a real-time noise detection module. Zhang's study used a model consisting of multiple layers of convolutional blocks and transition blocks. The transition blocks are used so that the features are reduced by half. This method resulted in F1-score of 0.8150, Cohen's Kappa(K) of 0.7276 and accuracy of 81% [1].

### 2.2 Neural Network Models for Sleep Stage Classification

Sun et al. [9] introduced a novel sleep stage classification method based on a Window Deep Belief Network (WDBN) stage and a Long Short-Term Memory (LSTM) stage to enhance the F1-score of classification to 0.806. The WDBN used in this method can extract features from EEG signal and combine them with the hand-crafted features to enhance the accuracy of sleep stage classification. However, this model is heavily dependent on the prior knowledge to extract hand-crafted features. Furthermore, Kang et al. [10] proposed a new 3 layers LSTM model with batch-normalisation to enhance the accuracy of multi-class sleep event classification. This model used batch normalisation to eliminate inconsistencies present in the output variance. However, LSTM with 4 layers is prone to overfitting. As a solution to the overfitting problem, this paper uses dropout technique to prevent overfitting.

Sors et al. [6] proposed a method for sleep stage scoring using temporal context and feature extraction capabilities of CNN. This work was done to enhance the performance and to make the method available on portable and wearable devices. In addition, this method can be used in continuous monitoring of brain injury patients. This paper used a 1-Dimensional Convolutional Neural Network (1D CNN) model with temporal context that led to the study of Zhang et al. [3], and Fernandez-Blanco et al. [2] to develop end-to-end network training. This method achieved an accuracy of 87% and Cohen's Kappa(K) of 0.81. Furthermore, Fernandez-Blanco et al. [2] proposed an architecture that makes use of simultaneously recorded EEG signals to test the system separately with each signal and also as a combined input. As a result, the two signal inputs were found to be the most advantageous with an accuracy of 92.67% and Cohen's Kappa(K) of 0.84 on two channels. Werth et al. [11] proposed a model that improves the accuracy of sleep stage classification in infants by using an ECG R-peak detection algorithm. This work uses Adam optimiser to improve the learning rate and remove the learning rate decay in the model. It has been shown that Adam optimiser is the most efficient and superior performing optimization algorithms.

Erdenebayar et al. [12] compared the accuracy of six different neural network models and identified the most accurate method to classify events of sleep stage. As a result, the best performing model, 1D CNN, showed an accuracy of 99% in detecting class events. Mousavi et al. [5] proposed a method to enhance the accuracy and Cohen's Kappa coefficient of sleep stage classification using data augmentation on a raw EEG signal. This method eliminated the need of a feature extraction stage as the model is capable of learning proper features related to each case. This method non-linearizes the network by using a ReLU activation function as a result of which the problem of gradient saturation is avoided. The simulated model resulted in accuracy for 2 to 6 stage classifications as 92.95% to 98.10%.

Zhang et al. [3] enhanced the CNN method by introducing orthogonal weight initialisation. This model used time-frequency image representation to learn rich and diverse features of the input signal. This is done by using Hilbert-Huang Transform to convert 1-Dimensional signal into 2-Dimensional time-frequency representation, this resulted in accuracy of 88.4. This work is further discussed in the below section.

## 2.3 State of Art

This section presents the current system features (highlighted in blue broken lines in Figure 2 and the limitations (highlighted in red broken lines in Figure 2). Zhang et al.[3] proposed an OCNN system to be used instead of traditional CNN system in order to better learning rich and diverse features of the EEG signal. This system makes use of the Hilbert Huang transform that converts EEG data to a Time Frequency Image representation which better describes the EEG signal for a CNN. This paper conducted the research by fusing an autoencoder and the Squeeze and Excitation network into one block to improve the performance of the classification algorithm and dramatically decrease the training time. It provides an accuracy of 88.4%. This model consists of three stages as shown in Figure 2., i.e. Pre-processing, Time Frequency Image Representation, and Orthogonal Convolutional Neural Network [3].

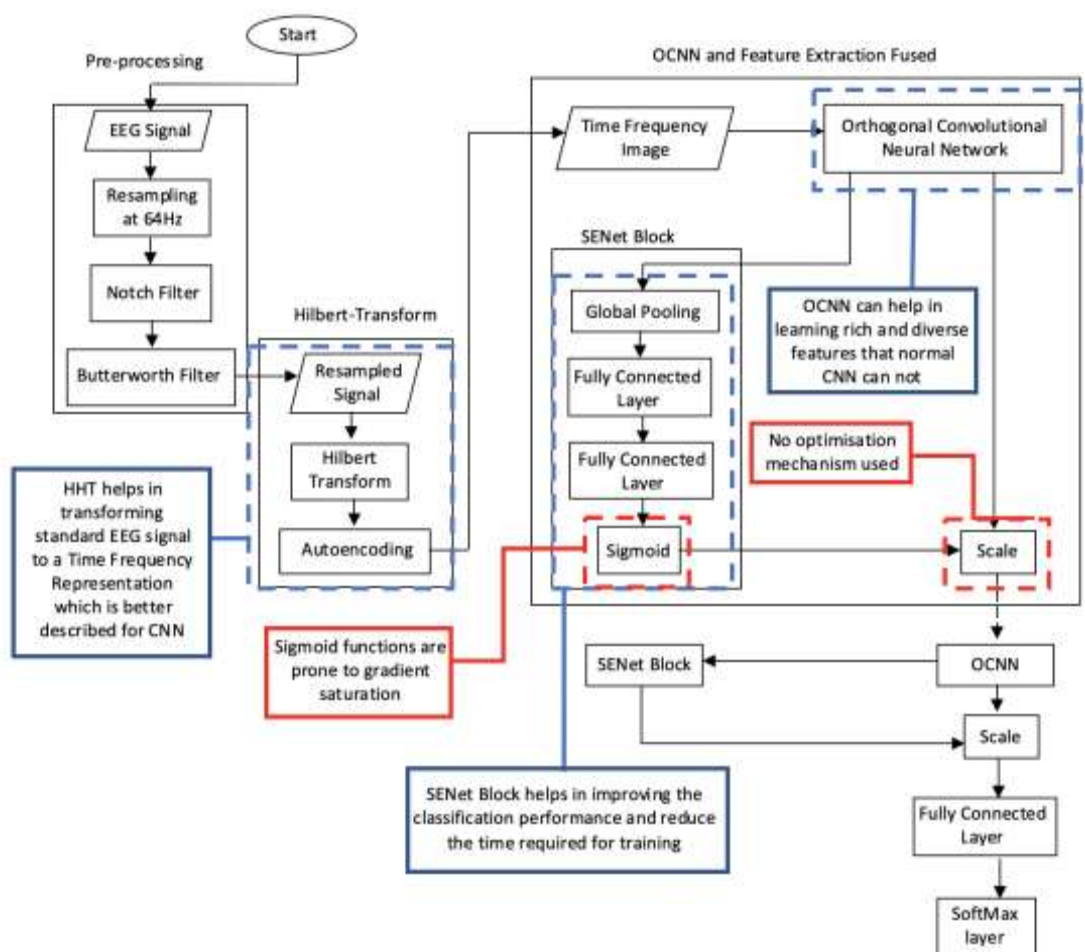

Figure 2 - Block Diagram of the State of Art Model.
[The blue borders show the good features of this state of art solution, and the red border refers to the limitations of it]

**Pre-processing:** Pre-processing stage starts with resampling of the EEG signals at 64Hz to remove the resampling differences in different datasets. Notch filtering is applied at 50/60Hz so that the power line differences are cancelled out. Then to filter out the resampled signals to 0.5 – 30 Hz, an eighth order Butterworth filter is used.

**Time Frequency Image Representation:** Zhang et al. [3] used time frequency image representation instead of EEG as an input to the CNN to better learn rich and diverse features. To achieve this, the first step is empirical mode decomposition, and then Hilbert Transform operate on the result to construct a time frequency signal. Eventually, the output is run through the autoencoder to reduce the dimension of the signal.

**Orthogonal Convolutional Neural Network:** In this stage, deep weights are initialized to improve the learning speed. After that orthogonal regularization is performed. Then the Squeeze and excitation block is implemented to improve the classification performance and reduce the training time. The implementation of the SENet block helps increase the weight of useful features and decrease the weight of useless features.

**Limitation:** However, this model used a sigmoid function as the activation function.  due to which the linear part of each neuron will have a very big or very small amount of value after some epochs of training. Since the gradient values become very small, weights are updated very slowly. To mitigate this, ReLU function needs to be used. This will directly result in an improvement in the classification accuracy as the weights are updated faster. With this model, there is no optimisation involved. It is always beneficial to add an optimisation algorithm for more balanced classification and better training. The use of right optimisation algorithm will result in a reduction of the learning time, i.e. significant improvement in the processing time.

**Limitation Justification:** With the use of a ReLU function instead of a sigmoid function, the problem of gradient saturation can be removed and eventually, the classification accuracy can be increased. With the use of an Adam optimizer, the training time can be decreased as it can help to reduce the noise on the training signal data, which is beneficial for the portable sleep stage detection devices. This model presented a classification accuracy of 88.4% and 87.6% on two public datasets, respectively.

The OCNN Algorithm is implemented with 7 convolution layers, one fully connected layer and one SoftMax layer. Since this model also uses sigmoid as the activation function, gradient values become very small. The weights are updated very slowly, and this causes gradient saturation. To mitigate the gradient saturation, ReLU function can be used. This will directly result in an improvement in the classification accuracy as the weights are updated faster. Also, the processing time can be reduced using an Adam optimiser.

The sigmoid function is given by Equation (1) [3]:

$$Y(y) = \frac{1}{1 + e^{-y}} \qquad (1)$$

Where,
e = the natural logarithm
y = the output feature map can be expressed by Equation (2) [3]:

$$y_i = f\left(\sum_{i=1}^{N} W_{i,j} \cdot X_i + b\right) \qquad (2)$$

Where,

$y_i$ = Output activation.

$W_{i,j}$ = The weights, as explained in Equation (2a).

$X_i$ = Input activation.

b = The bias term.

N = Number of inputs to the neuron.

$$W^* = \begin{bmatrix} W^1 \\ W^2 \\ \vdots \\ W^N \end{bmatrix} = \begin{bmatrix} W_1^1 W_2^1 \dots W_N^1 \\ W_1^2 W_2^2 \dots W_N^2 \\ \vdots \\ W_1^N W_2^N \dots W_N^N \end{bmatrix} \qquad (2a)$$

Where,

$W^*$ = Is an orthogonal matrix, used to store the initialised weights for each layer.

The Convolution model implemented with 7 convolutional layers does not optimise the weights before initialising them. By using an Adam optimiser, the weights are optimised before they are initialised to the system which results in a reduced training time and processing time. Table 1 presents the algorithm of the state of art OCNN solution, while Figure 3 shows the flowchart of this solution.

Table 1 - Algorithm of the state of art OCNN model.

| Algorithm: Orthogonal Convolutional Neural Network Input: Time frequency Image representation of EEG signal after Hilbert Huang Transform |
| --- |
| Output: Sleep Stage Class probability |
| BEGIN |
| Step 1: Get the time frequency image representation of the signal from the pre-processing module. |
| Step 2: Feed the signal into the orthogonal neural network model of 7 layers. |
| Step 3: Send two copies of the output of OCNN, one to the SENet block and one to the combination block. |
| Step 4: If signal is received into the SENet block, feed it to the global pooling block and fully connected layers, else go to step 6. |
| Step 5: Feed the signal into the Sigmoid operator as in equation (1) |
| Step 6: Combine the signals from SENet Block. |
| Step 7: Feed the output to Next layer in the OCNN model and repeat Step 3 to 7, 6 more times. |
| Step 8: Feed into the fully connected layer.. |
| Step 9: Feed into the SoftMax Layer |
| END |

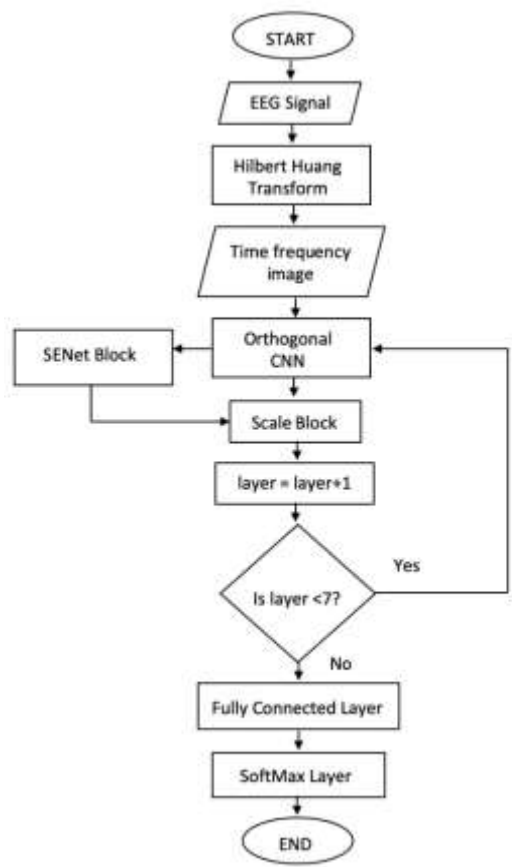

Figure 3 - Flowchart of OCNN algorithm

## 3. The Proposed System

After reviewing different literatures regarding the sleep stage classification algorithm, and analysing the pros and cons of each method, it was found that the OCNN algorithm proposed by Zhang et al.[3], is the most effective algorithm to determine the rich and diverse features of an EEG signal. These features make use of orthogonal initialisation and orthogonal regularisation to significantly improve the learning speed of the training models. This paper also proposes a method to convert the EEG to a time frequency domain by using Hilbert-Huang Transform function. These methods improve the overall accuracy of the system and reduce the learning time. The OCNN algorithm makes it possible for the model to run on portable sleep devices as they are not complicated as the normal CNN methods.

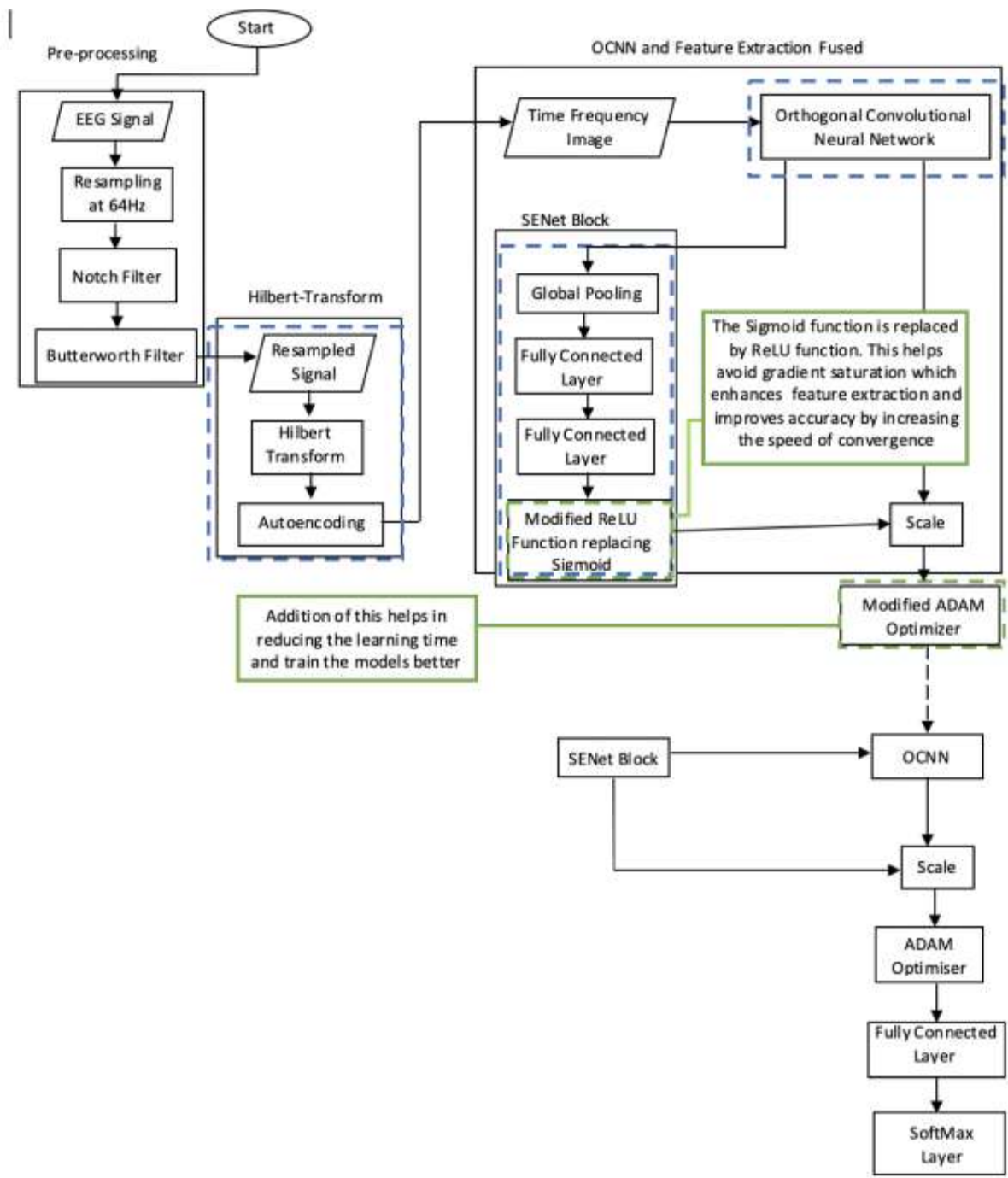

Figure 4 - Block diagram of the proposed OCNN model with Adam optimiser
[The green borders refers to the new parts in our proposed system]

The proposed solution with Adam optimizer will help reduce the processing time by reducing noisy parameters. This feature has been adapted from the second-best solution [2]. It is always beneficial to add optimisation algorithm for more balanced classification and better training of the models. The proposed system consists of three major stages, as shown in Figure 4, called Pre-processing, Time Frequency Image Representation, and Orthogonal Convolutional Neural Network. These stages are explained with more details in the following paragraphs:

**Pre-processing:** This stage removes the resampling differences in the different datasets which are used in the system. The signals are first resampled at 64Hz, then it is passed through a Notch filter at 50/60Hz so that the power line differences are cancelled out. Then to filter out the resampled signals to 0.5 – 30 Hz, an eighth order Butterworth filter is used. The datasets are oversampled to eliminate the class imbalance in the sleep stages. This data is then fed into the next component.

**Time Frequency Image Representation:** This stage is one of the most significant features of the proposed model. First step is empirical mode decomposition, and Hilbert Transform operate on the result to construct a time frequency signal. Then the output is run through the autoencoder to reduce the dimension of the signal.

**Orthogonal Convolutional Neural Network:** In this stage, ReLU function is used instead of the sigmoid function and this directly result in an improvement in classification accuracy as the weights are updated faster. With the addition of Adam optimizer, the proposed models are better trained with reduced time frames. ReLU function and Adam optimized remove the limitation of the state of art as described in section 2.3.

### 3.1 Proposed Equation

Equation 3 represents the ReLU function. It has been used as the activation function for the neural network model in the second-best solution [5]. We replaced the sigmoid activation function used by the state of art by the ReLU function as it will get rid of the limitation of gradient saturation that occurs in sigmoid function. A ReLU activation function will ensure that gradient does not vanish, and the weights are updated faster. The removal of gradient saturation will contribute to improvement in accuracy of the sleep stage classification.

$$f(x) = \begin{cases} x & if\ x > 0 \\ 0 & otherwise \end{cases} \tag{3}$$

Equation (4) represents leaky ReLU function [6]. ReLU functions can have a problem of dying ReLU, it returns a zero for all negative values. There is a tendency for neuron to vanish if it has a negative value [13]. The Leaky ReLU function has a small slope to the negative values rather than a zero value. Equation (3) has been modified to Equation (4). This adds the negative value as a parameter to the model and thus reduces the training time by producing more balanced results.

$$f(x) = \begin{cases} x & if\ x > 0 \\ \alpha\ x & otherwise \end{cases} \tag{4}$$

Where,
$\alpha$ = A constant gradient.

Equation (5) represents Rectified Linear Unit (ReLU) function modified to include a slope of 0.1 for negative values. This allows the function to add negative value as a parameter to produce balanced results and improve the accuracy. We experimented with different values of α (0.1, 0.2, and 0.3) and ended up with 0.1 because it works best for our system, as shown in our results given in section 4.

$$Mf(x) = \begin{cases} x & if\ x > 0 \\ 0.1x & if\ x \leq 0 \end{cases} \quad (5)$$

We use this modified function as an activation function in equation (6) where the sigmoid function in equation (1) is replaced by a ReLU function in equation (6).

$$Y = Mf(x) \quad (6)$$

Equation (7) represents Adam optimisation algorithm [14], it is used to optimise each weight $W_i$ before initialisation. This is done to update the initial weight $w_i$ by optimising it to $W_i$ which represents the new optimised weight. This step will allow the model to learn useful features faster.

$$W_i = w_i - \alpha.dw_i \quad (7)$$

Equation (7a) gives the stochastic gradient descent that is calculated using the values of weighted average derivatives.

$$dw_i = \frac{vdw}{\sqrt{sdw}} \quad (7a)$$

Equation (7b) and (7c) are used to calculate the values of vdw and sdw that represent the exponentially weighted average of derivatives.

$$vdw = \frac{vdw_i}{1 - \beta 1} \quad (7b)$$

$$sdw = \frac{sdw_i}{1 - \beta 2} \quad (7c)$$

Equation (7d) represents the momentum for each weight initialisation with the hyperparameter represented as $\beta 1$ and $\beta 2$.

$$vdw_i = \beta 1.vdw_i + (1 - \beta 1).w_i \quad (7d)$$

Equation (7e) represents the RMSprop for each weight initialisation with the hyperparameter represented as $\beta 1$ and $\beta 2$.

$$sdw_i = \beta 2 . sdw_i + (1 - \beta 2) . (w_i)^2 \tag{7e}$$

Equation (8) represents the modified Adam optimiser with learning rate lr = 3x10$^{-5}$, $\beta_1$ = 0.9, $\beta_2$ = 0.999 [6]. The value for dw$_i$ is calculated from equation (7a). Equation (7) has been modified to equation (8) to incorporate the parameters on [6]. This will allow the proposed system to operate on recommended $\beta_1$ and $\beta_2$ values along with the learning rate of 3x10$^{-5}$

$$MW_i = w_i - lr . dw_i \tag{8}$$

The modified output feature map $My_{1,1}^1$ is given by equation (9). Equation (2) has been modified by us to include the modified weight initialisations derived from equation (8). This has been done to update the weights at each stage and optimise their values using the Adam optimiser. As a result, the training time is reduced.

$$My_{1,1}^1 = f\left(\sum_{i=1}^{N} X_{i,j} * MW_i^1 + b_1\right) \tag{9}$$

Enhanced Sleep Stage Classification is represented by equation (10). It consists of MY which represents the modified activation function Leaky ReLU as represented in equation (6). It also consists of a modified output feature map as represented in equation (9)

**ESSC** = $Y + My_{1,1}^1$ (10)

$Y$ = Modified activation function Leaky ReLU: Y= Mf(x).

$My_{1,1}^1$ = Modified Output feature map

Two equations are proposed, first is the ReLU function which aims to avoid the gradient saturation problem that is faced by most sigmoid functions. As the sigmoid function in the state of art is replaced by a ReLU function, the gradient saturation problem does not exist anymore, and the convergence speed is increased along with better classification accuracy.

The second proposal is an addition to output feature map, which introduces optimisation to reduce the weight of useless features and amplify that of the useful features which leads to decreased learning time and increased accuracy and performance.

For the experiment, 2.2 GHz Intel Core i7 processor with 8 GB RAM and a Nvidia RTX 2070 GPU is used. The deep learning framework TensorFlow has been used with Python and Keras framework to implement this solution. Evaluate () method of the Keras Python package is used to calculate the accuracy. The fit generator () method of Keras package is used to calculate processing time. This is done by the subtraction of start time and end time is calculated to get the execution time of the system.

**3.2 Databases**

Six main sleep databases are used for training and testing the proposed method, most of these sleep databases are free to download from PhysioNet [2]. There databases are St. Vincent University Hospital/ University College Dublin database (UCD), MIT-BIH Polysomnographic database (MIT-BIH) , Sleep EDF database, Sleep EDF Extended database, Montreal Archive of Sleep Studies database (MASS ), and Sleep Heart Health Study database (SHHS).

The UCD database includes 25 overnight polysomnograms collected from 25 adult subjects who are suspected of having sleep apnea. The subjects are monitored at St Vincent's University Hospital, Dublin. Each record consists of two EEG channels with sampling rate equal to 128 Hz, two EOG channels and one chin EMG channel with sampling rate equal to 64 Hz. Subject were randomly selected over six months from patients for possible diagnosis of obstructive sleep apnea, central sleep apnea, or primary snoring [3]. The MIT-BIH database contains records of multiple physiologic signals for patients during their sleep. The data was collected in Boston's Beth Israel Hospital Sleep Laboratory for evaluation of chronic obstructive sleep apnea. This database contains 16 recording from 16 male subjects with suspected chronic OSA. The recording contains ECG, EEG, and respiration signals with 250 Hz sampling rate [3]. The EDF database is a simple database used to store multichannel biological signals. This dataset has been collected from the body-worn bio signal amplifiers. The recording was collected from 26 subjects aged 21-35 years. The subjects are Caucasian females and males (selected randomly) without any medication. The recordings contain EOG and EEG with sampling rate equal to 100 Hz. The Sleep EDF database was a small subset contributed in 2002, It was greatly expanded in 2013 to contain 61 whole-night PolySomnoGraphic (PSG) sleep recordings with accompanying hypnograms. In 2018 the database was greatly expanded and now it contains 197 whole night PSGs sleep recordings with accompanying hypnograms. The sleep EDF Extended database containing EEG, EOG, chin EMG, and event markers. The Sleep EDF extended database helps to standardize the way in which the signals are used [15][16]. The MASS database is a collaborative dataset of laboratory-based PSG recordings. O'Reilly et al. [17] described this database. The recordings were organized into five sub datasets according to the research protocols from which they were collected to provide homogeneous cohorts This database comprises polysomnograms of 200 nights recorded in 97 men and 103 women. The ages of the subjects are between 18 and 76 years. The SHHS database contains data related to a prospective cohort study. The purpose of this study is to investigate the relationship between sleep apnea and cardiovascular disease. The SHHS dataset contains two sub datasets, SHHS1 and SHHS2. Each sub database contains one night of sleep EEG data for thousands of subjects collected between 1995 and June 2003. Random subjects were selected from SHHS1 and SHHS2 for both the training and testing process [15][16][17].

The Keras library first converts the edf data into a python readable array format, i.e. .npz that is a numpy gzip compression in the TensorFlow platform. Then we randomize the dataset into training and testing data. K fold cross validation was used with the value of k as 20. The sample data is randomly divided into k subsamples where one is used for testing and remaining for training. 15% of the data has been set aside for testing, and the remaining are used for training. After this, the 1-dimensional CNN model is initialized with orthogonal weights which are used to train the model. The trained model is stored as a .h5 file type which is a Hierarchical Data Format (HDF). The samples were scored manually according to the Rechtschaffen and Kales [18]. The accuracy is calculated as given in Equation 11 [3]:

$$\text{Accuracy} = \frac{True\ Positive + True\ Negative}{True\ Positive + False\ Negative + False\ Positive + True\ Negative}\% \qquad (11)$$

The formula for the standard deviation is

$$\sigma = \sqrt{\frac{\sum |x - \bar{X}|^2}{n}} \qquad (12)$$

σ = standard deviation

x = sample

$\bar{X}$ = mean of the sample

n = total number of samples

### 3.3 Area of Improvement

We proposed a Leaky ReLU function as an activation function along with a orthogonal weight initialisation method to avoid gradient saturation and to speed up the convergence to accurate result. This also helps to improve the classification accuracy. A sigmoid function is prone to reach its peak maximum or minimum value which leads to gradient saturation. This saturation stops the gradient propagation and results in improper weighting of features. Firstly, a ReLU function helps to avoid gradient saturation and to converge faster. This also helps improve classification accuracy. Secondly, the Adam optimizer allows the models to better train with reduced time frame which helps to improve accuracy and reduce training time. Adam optimisation achieves this by faster convergence and reduced oscillations increase momentum towards the optimal result.

The state of art uses sigmoid as an activation function, that has been replaced by a Leaky ReLU function in the proposed system to improve the accuracy of classification of sleep stages. Optimisation algorithm can help in minimizing the error function that depends on the model's learnable parameters. This step of optimisation is important to decrease the training time and improve the accuracy. The state of art system does not provide any sort of optimisation in training the model. Table 2 presents the algorithm of the proposed system, while Figure 5 shows the flowchart of the proposed OCNN algorithm.

Table 2 - Algorithm for the proposed system.

| Algorithm: Orthogonal Convolutional Neural Network with Leaky ReLU and Adam optimiser to reduce training time and improve accuracy |
| --- |
| Input: Time frequency Image representation of EEG signal after Hilbert Huang Transform |
| Output: Sleep Stage Class probability |
| BEGIN |
| Step 1: Get the time frequency image representation of the signal from the pre-processing module. |
| Step 2: Feed the signal into the orthogonal neural network model of 7 layers. |
| Step 3: Send two copies of the output of OCNN, one to the SENet block and one to the combination block. |
| Step 4: If signal is received into the SENet block, feed it to the global pooling block and fully connected layers, else go to step 6. |
| Step 5: Feed the signal into the ReLU block using equation (3). |
| Step 6: Combine the signals from SENet Block and OCNN block and feed into Adam optimiser. |
| Step 7: Feed the output to Next layer in the OCNN model and repeat Step 2 to 7, 6 more times. |
| Step 8: Feed into the fully connected layer. |
| Step 9: Feed into the SoftMax Layer. |
| END |

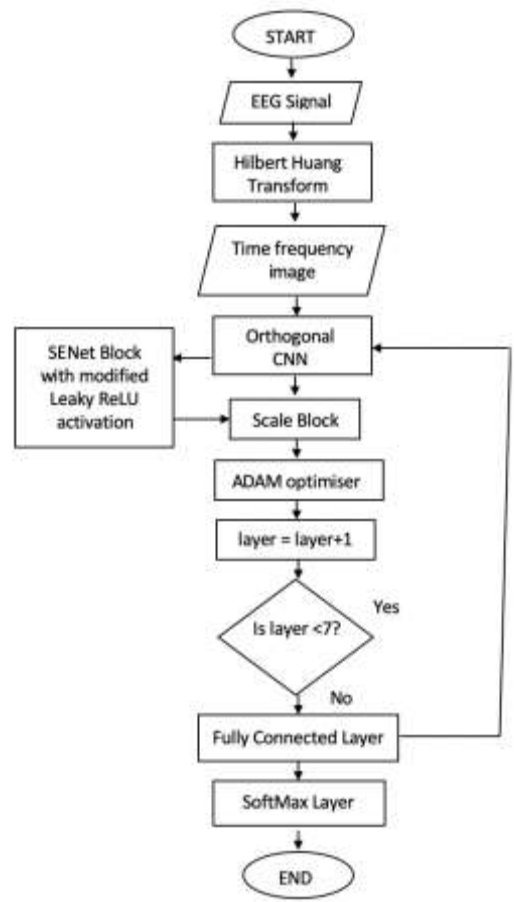

Figure 5 Flowchart of OCNN algorithm with Leaky ReLU activation and Adam optimisation

## 4. Results and Discussion

The results were compared with the state of art and other literatures based on accuracy, processing time, and F1 score. Table 3 shows a comparison of UCD dataset between the state of art and the proposed solutions.

Table 3 - Comparison of UCD Dataset (%).

| No. | Dataset | Sleep Stage | Classification Accuracy (%) | |
|---|---|---|---|---|
| 1 | UCD | | **State-of-art Solution** | **Proposed Solution** |
| | | Wake | 89 | 91 |
| | | S1 | 82 | 83 |
| | | S2 | 94.5 | 92 |
| | | SWS | 91.7 | 100 |
| | | REM | 86 | 89 |
| **Total** | | **Average Accuracy (%)** | **88.64** | **91** |
| | | **Processing Time** | **1250s** | **1192s** |

The sequence test for UCD dataset (given in Table 3) shows a 2.36% increase in accuracy. The change in classification accuracy for the classification stages Wake, S1, S2, SWS and REM are 2, 1, -2.5, 8.3 and 3 respectively. The accuracy for classification of S2 stage tends to fall but the accuracy of classification for Wake, S1, SWS and REM have shown significant increase. This might be related to the random selection of the subjects involved in the training and testing process. Using selective procedure, especially when creating the training set, might efficiently affect the accuracy. Furthermore, the accuracy for classification of SWS is too high (100%), the main reason for this excellent accuracy is related to the data used in the training and testing. Also, it is related to the types of subjects in the UCD database. All subjects were randomly selected from patients referred to the Sleep Disorder Clinic for possible diagnosis of apnea or primary snoring.

Table 4 shows the sequence test for MIT-BIH database, there is 7.2 increase in accuracy from 88.1. Table5 shows the sequence test for EDF dataset, the accuracy increased by 3.32 from 83.9. For EDF Expanded dataset given in Table 6, the accuracy is increased by 2.74 from 85.86. In addition, the processing times are decreased due to faster learning and prediction capability, as given in Table 3, Table 4, Table 5, and Table 6. For example, the processing time for UCD database is decreased by 58s to 1192s. For MIT-BIH, the processing time is decreased by 49s to 1191s. For Sleep EDF database, it is decreased by 87s to 1203s. Also, for EDF Expanded database, it is decreased by 72s to 1188s.

Table 4 - Comparison of MIT-BIH dataset (%).

| No. | Dataset | Sleep Stage | Classification Accuracy (%) | |
|---|---|---|---|---|
| 2 | MIT-BIH | | **State-of-art Solution** | **Proposed Solution** |
| | | Wake | 87 | 90 |
| | | S1 | 80.4 | 86 |
| | | S2 | 95.1 | 96 |
| | | SWS | 92.4 | 95 |
| | | REM | 85.6 | 81 |
| **Total** | | **Average Accuracy (%)** | **88.1** | **89.6** |
| | | **Processing Time** | **1240s** | **1191s** |

Table 5 - Comparison of Sleep EDF dataset (%).

| No. | Dataset | Sleep Stage | Classification Accuracy (%) | |
|---|---|---|---|---|
| 3 | Sleep EDF | | **State-of-art Solution** | **Proposed Solution** |
| | | Wake | 89 | 91 |
| | | S1 | 82 | 84 |
| | | S2 | 87.5 | 89 |
| | | SWS | 83 | 90 |
| | | REM | 78 | 82 |
| **Total** | | **Average Accuracy (%)** | **83.9** | **87.2** |
| | | **Processing Time** | **1290s** | **1203s** |

Table 6 - Comparison of Sleep EDF Expanded dataset (%).

| No. | Dataset | Sleep Stage | Classification Accuracy (%) | |
|---|---|---|---|---|
| 3 | Sleep EDF Expanded | | **State-of-art Solution** | **Proposed Solution** |
| | | Wake | 90 | 93 |
| | | S1 | 83 | 85 |
| | | S2 | 88.3 | 89 |
| | | SWS | 86 | 92 |
| | | REM | 82 | 84 |
| **Total** | | **Average Accuracy (%)** | **85.86** | **88.6** |
| | | **Processing Time** | **1260s** | **1188s** |

An improvement in the accuracy has been seen as the model is trained with more datasets, as shown in Table3, Tables 4, Table 5, and Table 6. The classification results are compared with the manually labelled classification data to calculate the accuracy for Wake, S1, S2, SWS and REM stages. The processing time is compared from the output table for each dataset. It shows a steady decrease in processing time for each dataset. The accuracy is calculated as given in Equation 11, and the mean and standard deviation of the sample data of four different stages of sleep are calculated using Equation 12.

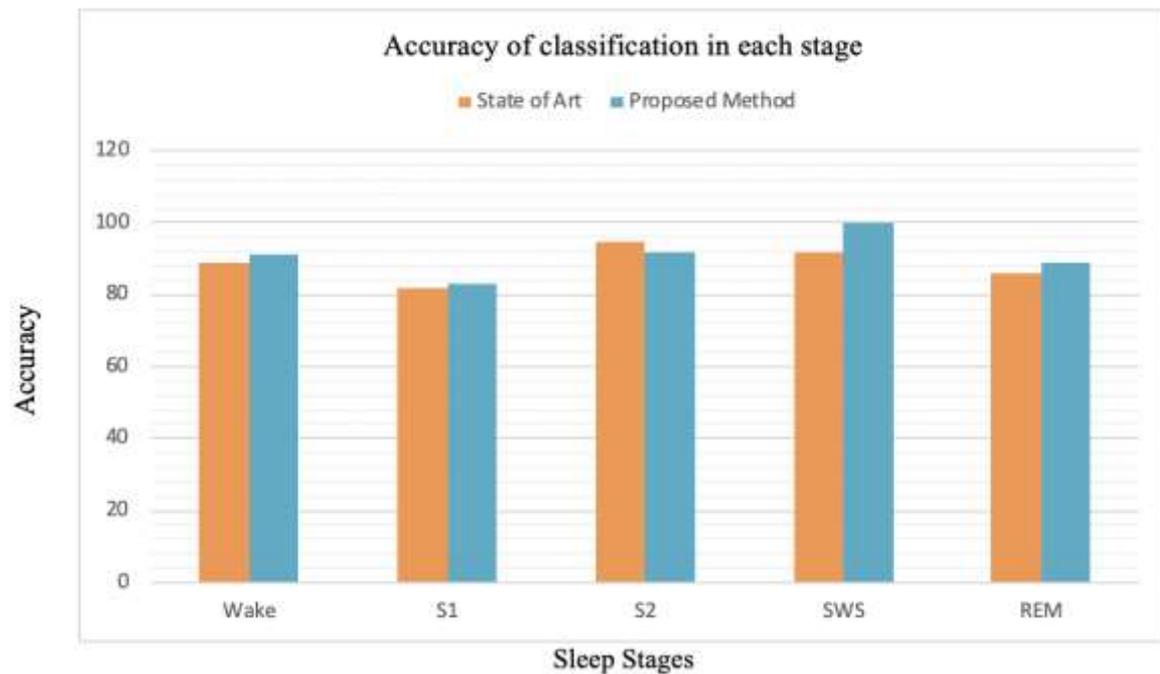

Figure 6 - Accuracy of classification in state of art and proposed method for each sleep stage

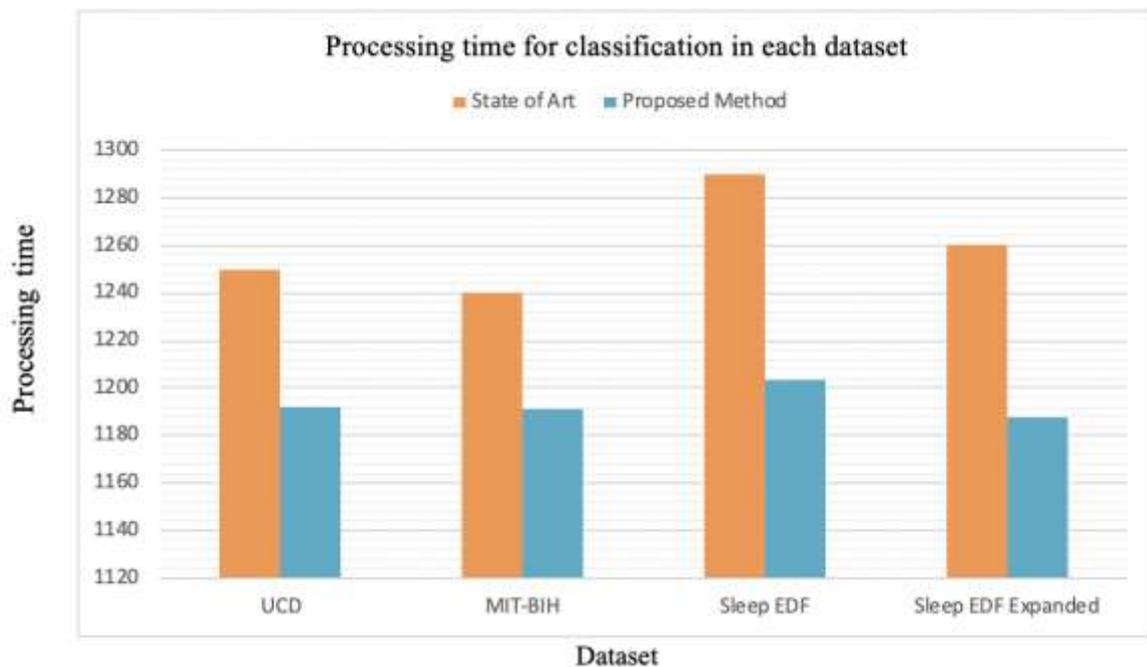

Figure 7 - Processing time of classification in state of art and proposed method for each sleep stage

Figure 6 shows a comparison of accuracy of classification between the state of art and the proposed method for each stage of sleep i.e. Wake, S1, S2, SWS, REM. The proposed method (blue color) has shown an increased overall the state of art method (the orange color) for all sleep stage. The figure 7 shows a comparison of processing time for classification between state of art and proposed method for each database i.e. UCD, MIT-BIH, Sleep EDF, and Sleep

EDF Expanded. The proposed solution has shown an average decrease in the processing speed by 58 second for UCD dataset, 49 second for MIT-BIH, 87 second for EDF, and 72 second for EDF Expended. The proportion of true positive results is calculated to obtain the accuracy where true positive and true negative metrics are used.

The results of our proposed system given in Table 3, Table 4, Table5, Table 6 show that the improvement in the classification accuracy of S1 and REM were lower than the improvement of the other stages like Wake, S2, and SWS. In table 3, for UCD database, although the accuracy of classification for Wake, S1, SWS and REM have shown significant increase, however, the accuracy for classification of S2 stage tends to fall. Thus, the diverse and features of S1 and REM require further research and improvement. Furthermore, currently, we randomize the dataset into training and testing data. However, using selective procedure specially when creating the training set, might affect the classification accuracy of our proposed method.

The improvement in accuracy is a contribution of the activation function ReLU that has reduced the problem of gradient saturation in state of art. Furthermore, the reduction in processing time has been achieved through the use of optimised weights that has significantly reduced the amount of time taken to reach to the optimal result for each neuron. Table 7 summaries those results and gives a comparison between the state of art and the proposed solutions. From Table 7, the proposed system increases in accuracy due to faster and richer feature extraction than the state of art systems. Also, the proposed system decreases in processing time due to faster learning and prediction capability. Thus, our Enhanced Sleep Stage Classification system can be used for wearable sleep devices, such device can be used for home sleep monitor.

Table 7 - Comparison Table

|  | **Proposed Solution** | **State of Art Solution** |
|---|---|---|
| **Name of the solution** | Leaky ReLU activation function and Adam optimisation | Uses Sigmoid as an activation function and no optimisation technique |
| **Accuracy** | Increase in accuracy due to faster and richer feature extraction. The accuracy increased as following:<br>• UCD, by 2.36%, from 88.63 to 91%.<br>• MIT-BIH, by 1.5, from 88.1 to 89.6<br>• EDF, by 3.3, from 83.9 to 87.2<br>• EDF Expanded, by 2.74, from 85.86 to 88.6. | Comparatively lower accuracy:<br>• UCD, by -2.36%, from 88.64 to 91%.<br>• MIT-BIH, by -1.5, from 88.1 to 89.6<br>• EDF, by -3.3, from 83.9 to 87.2<br>• EDF Expanded, by -2.74, from 85.86 to 88.6. |
| **Processing/Training time** | Decrease in processing time due to faster learning and prediction capability. The processing time decreased by:<br>• UCD, by 58s to 1192s.<br>• MIT-BIH, by 49s to 1191s.<br>• EDF, by 87s to 1203s.<br>• EDF Expanded, by 72s to 1188s | Larger training time due to no optimisation. The training time is approximately:<br>• UCD, 1250s.<br>• MIT-BIH, 1240s.<br>• EDF, 1290s.<br>• EDF Expanded, 1260s. |
| **Contribution 1** | Replacement of the sigmoid function by a Leaky ReLU function to avoid gradient saturation problem and to increase convergence speed and accuracy.<br><br>$$MY = \begin{cases} y & if\ y > 0 \\ 0.1y & if\ y \leq 0 \end{cases} \quad (5)$$ | The state of art uses a sigmoid function which is prone to suffer from gradient saturation.<br><br>$$Y = \frac{1}{1+e^{-y}} \quad (1)$$ |

| **Contribution 2** | Addition of an Adam optimiser helps achieve faster convergence towards useful classification. This helps reduce noise which in turn can result in reduction of training time. $$W^* = \begin{bmatrix} W^1 \\ W^2 \\ \vdots \\ W^N \end{bmatrix} = \begin{bmatrix} W_1^1 W_2^1 \dots W_N^1 \\ W_1^2 W_2^2 \dots W_N^2 \\ \vdots \\ W_1^N W_2^N \dots W_N^N \end{bmatrix} \quad (2a)$$ | The state of art does not provide a method for optimisation of signals. $$W_i = w_i - \alpha.dw_i \quad (7)$$ Where $w_i$ represents each initialised weight $\alpha$ represents the step size |
|---|---|---|

# 6. Conclusion and Future Work

The proposed solution aims to improve the sleep stage classification accuracy by removing the problem of gradient saturation. Replacement of the sigmoid function by a Leaky ReLU function can avoid the gradient saturation problem and increase convergence speed and accuracy. Along with this, the proposed solution aims to use optimised values as feature weights to reduce the processing time. Adam optimiser helps to reduce noise which in turn can result in reduction of training time. The proposed model shows an overall increase in accuracy and reduction in processing time. The limitation of the State of Art has been solved in this research by using Re. The comparison of UCD dataset between the state of art and our proposed method shows a 2.36% increase in accuracy and 58s decrease in the processing time. For MIT-BIH database, there is 7.2% increase in accuracy and 49s decrease in the processing time. For EDF dataset, the accuracy increased by 3.32% from and the processing time is decreased by 87s. Also, for EDF Expanded dataset, the accuracy is increased by 2.74%, and the processing time is decreased by 72 second compared with the state of art solution. The proportion of true positive results is calculated to obtain the accuracy where true positive and true negative metrics are used. In addition, it can be concluded that the best performing models constituted of 1-Dimensional CNN models.

Future research includes the application of advanced CNN models that will allow extraction of richer and more diverse characteristics from the raw signal to further improve the accuracy of the system. A further field of study is in the direction of maintenance of dynamic isometry with the use of ReLU functions.

To improve our results and the learning process in general, more databases can be used, for example, the ISRUC-Sleep dataset. In addition, currently, we randomize the dataset into training and testing data. However, using selective procedure specially when creating the training set, might efficiently affect the learning and the generalization abilities of our proposed method.

**Abbreviations Table**

| CNN | Convolutional Neural Network |
|---|---|
| RNN | Recurrent Neural Network |

| LSTM | Long Short Term Memory |
|------|------------------------|
| FFT  | Fast Fourier Transform |
| STFT | Short-Time Fourier Transform |
| HHT  | Hilbert Huang Transform |
| EEG  | Electroencephalogram |
| ECG  | Electrocardiogram |
| W    | Wake state |
| S1   | Sleep Stage 1 |
| S2   | Sleep Stage 2 |
| SWS  | Sleep Wake State |
| REM  | Rapid Eye Movement |
| ESSC | Enhanced Sleep Stage Classification |
| ReLU | Rectified Linear Unit |